\documentclass[conference]{IEEEtran}
\IEEEoverridecommandlockouts
\usepackage{cite}
\usepackage{amsmath,amssymb,amsfonts}
\usepackage{todonotes}
\usepackage{verbatim}
\usepackage{algorithmic}
\usepackage{graphicx}
\usepackage{textcomp}
\usepackage{xcolor}
\usepackage{footnote}
\usepackage{tikzpagenodes}

\makesavenoteenv{tabular}
\makesavenoteenv{table}
\def\BibTeX{{\rm B\kern-.05em{\sc i\kern-.025em b}\kern-.08em
    T\kern-.1667em\lower.7ex\hbox{E}\kern-.125emX}}

\begin{document}

\title{Towards an Edge Intelligence-based Traffic Monitoring System\\
%{\footnotesize \textsuperscript{*}Note: Sub-titles are not captured in Xplore and should not be used}
\thanks{We acknowledge the support of the PNRR project FAIR-Future AI Research (PE00000013), Spoke 9-Green-aware AI, under the NRRP MUR program funded by the NextGenerationEU and of the European Community's Horizon Europe Program under the MLSysOps Project, grant agreement 101092912}
}

\author{\IEEEauthorblockN{1\textsuperscript{st} Vincenzo Barbuto}
\IEEEauthorblockA{\textit{Dept. of Informatics, Modelling,} \\
\textit{Electronics and Systems (DIMES)} \\
\textit{University of Calabria}\\
Rende, Italy \\
vincenzo.barbuto@dimes.unical.it}
\and
\IEEEauthorblockN{2\textsuperscript{nd} Claudio Savaglio}
\IEEEauthorblockA{\textit{Dept. of Informatics, Modelling,} \\
\textit{Electronics and Systems (DIMES)} \\
\textit{University of Calabria}\\
Rende, Italy \\
csavaglio@dimes.unical.it}
\and
\IEEEauthorblockN{3\textsuperscript{rd} Roberto Minerva}
\IEEEauthorblockA{\textit{Dept. of Wireless Networks and} \\
\textit{ Multimedia Services} \\
\textit{Telecom SudParis, Institut}\\
\textit{ Polytechnique de Paris} \\
Evry, France \\
roberto.minerva@telecom-sudparis.eu}
\and
\IEEEauthorblockN{4\textsuperscript{th} Noel Crespi}
\IEEEauthorblockA{\textit{Dept. of Wireless Networks and} \\
\textit{ Multimedia Services} \\
\textit{Telecom SudParis, Institut}\\
\textit{Polytechnique de Paris} \\
Evry, France \\
noel.crespi@mines-telecom.fr}
\and
\IEEEauthorblockN{5\textsuperscript{th} Giancarlo Fortino}
\IEEEauthorblockA{\textit{Dept. of Informatics, Modelling,} \\
\textit{Electronics and Systems (DIMES)} \\
\textit{University of Calabria}\\
Rende, Italy \\
g.fortino@unical.it}
}

\maketitle

\begin{tikzpicture}[remember picture,overlay,shift={(current page.north)}]
\node[anchor=north,yshift=-0.3cm]{\includegraphics[width=17cm]{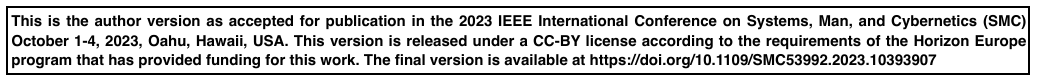}};
\end{tikzpicture}

\begin{abstract}
Cities have undergone significant changes due to the rapid increase in urban population, heightened demand for resources, and growing concerns over climate change. To address these challenges, digital transformation has become a necessity. Recent advancements in Artificial Intelligence (AI) and sensing techniques, such as synthetic sensing, can elevate Digital Twins (DTs) from digital copies of physical objects to effective and efficient platforms for data collection and in-situ processing. In such a scenario, this paper presents a comprehensive approach for developing a Traffic Monitoring System (TMS) based on Edge Intelligence (EI), specifically designed for smart cities. Our approach prioritizes the placement of intelligence as close as possible to data sources, and leverages an “opportunistic” interpretation of DT (ODT), resulting in a novel and interdisciplinary strategy to re-engineering large-scale distributed smart systems. The preliminary results of the proposed system have shown that moving computation to the edge of the network provides several benefits, including (i) enhanced inference performance, (ii) reduced bandwidth and power consumption, (iii) and decreased latencies with respect to the  classic cloud-centric approach.
\end{abstract}


\section{Introduction}

The United Nations' “The World's Cities in 2018” booklet predicts that by 2030, nearly 28\% of the world's population will reside in cities with at least 1 million inhabitants, and 8.8\% will live in megacities with over 10 million residents \cite{onu}, with relevant social and economic consequences. Therefore, motivated also by the urgent global issue of climate change, researchers have started seeking sustainable solutions for urban activities that have a significant impact on air pollution, such as the growing problem of traffic congestion \cite{goal11}. Addressing this issue is critical, as traffic bottlenecks not only result in higher carbon emissions, but also have a detrimental effect on the quality of life of urban residents.

In recent years, there has been a growing trend of utilizing the Internet of Things (IoT) \cite{atzori2010internet} in conjunction with wireless sensor networks (WSNs) \cite{lewis2004wireless} to develop sustainable solutions for complex systems, such as smart cities. One particular area where these technologies have demonstrated significant promise is in traffic monitoring and control applications. By deploying a network of sensors and actuators, these systems can collect and analyze accurate real-time data, which can aid in managing traffic flow and reducing congestion in urban areas \cite{thakur2016real}. Additionally, monitoring environmental parameters such as air quality and noise levels in real-time can enable authorities to make informed decisions about managing traffic and reducing negative environmental impacts. However, implementing IoT and WSNs necessitates an efficient and distributed software infrastructure capable of managing the enormous amounts of data generated and the Digital Twin (DT) \cite{minerva2020digital} represents a suitable enabling paradigm.  

By collecting and analyzing real-time data from sensors placed on physical objects (POs), DTs create virtual counterparts known as logical objects (LOs), which allow for more efficient decision-making and resource management. With its ability to monitor object life cycles, simulate behavior, and optimize performance, DTs have paved the way for real-time monitoring and optimization of various infrastructures and services through virtualization. Recent advances in distributed edge intelligence (EI) \cite{barbuto2023disclosing} have unlocked the full potential of DTs. This emerging paradigm, resulting from the convergence of edge computing and artificial intelligence (AI), equips edge devices with lightweight AI techniques. As a result, the DT concept becomes a promising platform not only for data collection, but also for infusing intelligence into intricate operating systems.

On this basis, within this paper, we propose a novel approach to develop large-scale distributed smart systems through the use of an EI-based Traffic Monitoring System (TMS) and the concept of the Opportunistic Digital Twin (ODT). In detail, in Section \ref{sec:relatedwork}, we provide a literature analysis on the current state-of-the-art of TMS with respect to related technology like cloud and edge computing and DT. In Section \ref{sec:proposal}, we present our proposal for an EI-based TMS, including its architecture, hardware, and software components. We also evaluate some preliminary results showing the benefits of using an EI-based approach rather than a cloud-centric solution in Section \ref{sec:results}. Finally, in Section \ref{sec:conclusion}, we conclude the paper by summarizing our findings and discussing some future implications.
\section{Related Work}
\label{sec:relatedwork}
DT, AI, and edge computing have never been simultaneously exploited in the context of a TMS. For instance, the use of DT technology is still in its initial phases and its precise role has yet to be defined, it is rapidly gaining popularity as a powerful tool for transportation management and planning. Andrey Rudskoy et al. \cite{RUDSKOY2021927}, analyze the implementation of an Intelligent Transport System (ITS) by providing a reference model of services which exploit machine learning techniques along with the new concept of DT. The goal was to reduce the human error inside traffic control centers by favoring operators activities automation. The proposed model shows that DTs could guarantee accurate representation of the real transport and road network for simulation purposes, to test several approaches based on predictive analytics and perform some calculation which aim to improve the whole transport system. Kumar et al. \cite{kumar2018novel} proposed a novel DT-centric approach for reducing traffic congestion by predicting driver intentions. Their method involved gathering real-time data from cameras and sensors along roads and bridges, using this data to construct virtual vehicle (VV) models as the DT of physical vehicles. The VVs, combined with historical driver data, were then used as inputs for machine learning and deep learning techniques to predict driver intentions. This approach also allowed VVs to interact with each other to predict other drivers' intentions, thereby paving the way for autonomous vehicles with path planning capabilities. Despite being novel and innovative, the aforementioned systems do not fully utilize the potential of IoT technology. Instead, they rely on expensive, specialized sensing and computing infrastructure because most of the machine learning solutions are executed in the distant servers and not in proximity of the data sources. Moreover, they consider DT to be only a high-fidelity modeling and simulation environment for real-world entities. While this statement is somewhat accurate, it is only a partial representation of the full scope of DT. In fact, Wang et al. argue in their recent publication \cite{wang2022mobility} that DT encompasses much more than just modeling and simulation. It also includes aspects such as physical sampling and actuation, as well as storage, modeling, learning, simulation, and prediction in the digital space. The authors' work introduces a novel Mobility DT (MDT) framework that employs AI to create a data-driven cloud-edge-device architecture for mobility services. The MDT framework is specifically designed to connect different mobility entities, with three physical building blocks: Human, Vehicle, and Traffic, each with an associated DT in the digital space. The authors emphasize the proposed hybrid architecture, as they understand that it is not always feasible to reach the cloud for executing complex machine learning tasks, particularly for autonomous vehicles that often lack internet connectivity. To address this challenge, they include (i) an edge layer capable of performing communication, caching, and heavy computing tasks offloaded by devices such as vehicles, and (ii) a device layer responsible for data acquisition, forwarding, and actuation processing. To demonstrate the feasibility of their proposed framework, the authors built an example cloud-edge architecture utilizing Amazon Web Services (AWS). This architecture accommodates the MDT framework and enables its digital functionalities of storage, modeling, learning, simulation, and prediction. The authors conducted a case study of the personalized adaptive cruise control (P-ACC) system, which integrates the key microservices of all three digital building blocks of the MDT framework. Nevertheless, the suggested use case remains highly dependent on cloud computing, with an edge layer able to carry out essential computing tasks, including raw data filtering and pre-processing.

It is worth noting that currently, intelligence mostly resides on cloud instances rather than being located as close as possible to data sources. Although the cloud is essential for large distributed systems, with several benefits like scalability and accessibility, relocating computation closer to data sources could have a significant impact on system performance. By doing so, organizations can reduce latency, enhance data security, and increase processing speeds, ultimately leading to more efficient and effective systems.
\section{Proposed Traffic Monitoring System}
\label{sec:proposal}
\begin{figure*}
    \centering
    \includegraphics[width=0.9\textwidth]{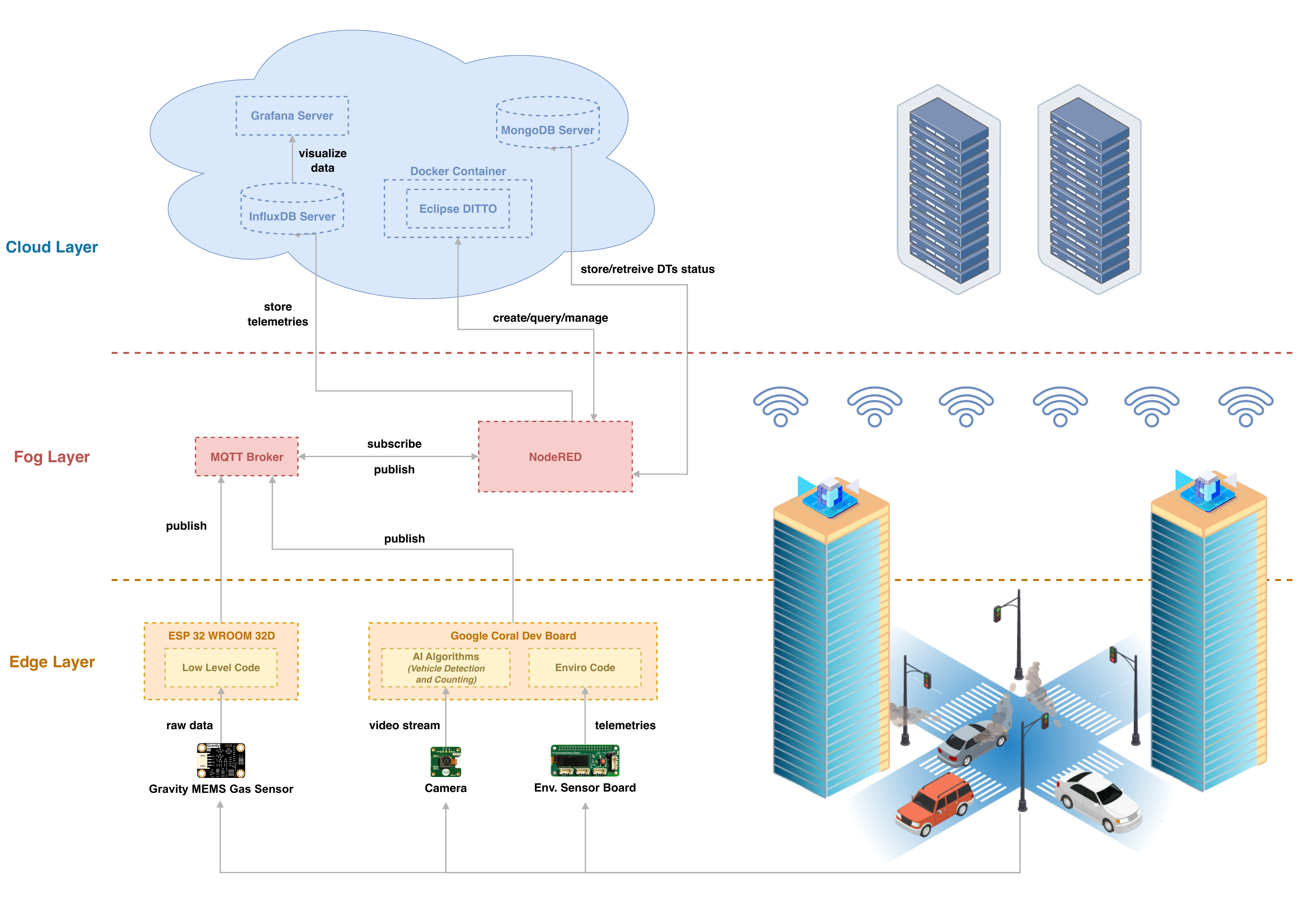
    }
    \caption{\label{fig:swoverview} Architecture of the EI-based TMS}
\end{figure*}
We propose an innovative TMS that leverages the EI paradigm and the concept of the ODT, which provides an efficient and effective structure for gathering and locally processing data generated by the sensors deployed along the target area. In particular, such sensory data from multiple sources are first stored within an ODT signature repository \cite{minerva2021exploiting} and later, through AI techniques, “opportunistically” selected and post-processed to generate new knowledge (e.g., estimate the traffic level by processing data about pollution and noise) which can be exploited by the TMS. 
With respect to conventional DT approach, the ODT improves modeling, reflection and entanglement between a tangible object and its digital counterpart.
%\textcolor{red}{The ODT concept, powered by EI, represents a significant advancement compared to traditional DT approaches. By enabling the selective extraction of relevant features and properties tailored to specific use cases and contextual requirements of both systems and physical environments, ODTs present a significant leap forward. ODTs employ a unique information model stored within a signature repository, enhancing their capacity for reflection and accuracy beyond that of classical DTs. Through generating more precise DTs, ODTs strengthen the bond or entanglement between systems or physical environments and their digital counterparts. This facilitates the efficient optimization and control of their behavior.}

To support the novel ODT approach, the proposed TMS incorporates the following features:
\begin{itemize}
    \item \textbf{EI-based technology} \cite{barbuto2023disclosing}: involves efficiently process data and executes AI algorithms such as vehicle detection and counting at the “edge” of the network. This approach significantly reduces processing times and minimizes the amount of data that needs to be transmitted over the network, resulting in increased efficiency and improved overall performance of the system. See Section \ref{sec:results} for details on the performance gains achieved with this technology.
    \item \textbf{Synthetic sensing over General-purpose devices} \cite{laput2017synthetic}: this approach involves using a single board to collect general measures related to both environmental and pollution data, and leveraging AI algorithms to generate additional insights (known as synthetic data) that traditional sensors cannot observe directly. By using this approach, we are able to monitor phenomena that would otherwise be impossible to observe with traditional sensors. This not only increases the system's capabilities, but also reduces deployment and maintenance costs, while making it more flexible and adaptable to different environments.
    \item \textbf{Data-driven and Bottom-up methodology} \cite{he2019data}: our approach prioritizes the use of data to inform key decisions and system design. Rather than relying on abstract models or assumptions as in the traditional top-down approach, we begin with specific sensory data and gradually construct a larger virtual structure. This approach enables us to monitor phenomena with greater accuracy and responsiveness.
\end{itemize}
Our chosen service deployment architecture is a hybrid-fog model, which involves distributing modules across multiple nodes throughout the network. Specifically, the edge layer encompasses all edge devices responsible for data gathering and real-time inference. Meanwhile, the fog layer houses dispatching and backend services, and the cloud layer contains servers for DT management, data persistence, and data visualization. The classic cloud-centric approach lends itself poorly to low throughput, high reliability connections. Using a fog architecture allows moving computation all over the network, (i) reducing communication latency and bandwidth consumption; (ii) improving the overall speed of the system. In the future, the proposed system will be tasked with analyzing a vast volume of data, including telemetries and video streams. Therefore, it is essential to design an architecture that can optimize data processing and display, as this will be crucial for the successful deployment of the final system.

\subsection{Use Case: Early Implementation}
To showcase the effectiveness of our approach, we started implementing the TMS subsystem aimed to collect real-time data on traffic, pollution, and meteorological conditions from the environment, process it using advanced deep learning techniques and low-latency, cost-effective communication technology, and store relevant information in a database for future analysis. Several hardware components are employed to achieve such a specific objective. The first component is a general-purpose board, which is the Google Coral Dev Board. It comes with a Tensor Processing Unit (TPU) capable of executing up to four trillion operations per second \cite{TPUperformances}, making it ideal for implementing advanced EI techniques directly on-device. Two sensor boards, the Gravity MEMS Gas Sensor Board, and the Google Environmental Sensor Board, collect data on pollutants and environmental telemetry. The camera module, either the Google Coral Camera or a USB camera, captures real-time video streams of vehicles on the monitored road. Finally, the microcontroller unit (MCU), the ESP32 WROOM 32D Board, interfaces with the Gas Sensor Board.

\begin{table}[h]
\centering
\caption{Breakdown of implementation technologies by field of application.}
\label{tab:purposes}
\resizebox{\columnwidth}{!}{%
\begin{tabular}{|c|c|c|}
\hline
\textbf{Purpose}                    & \textbf{Technology} & \textbf{Layer} \\ \hline
EI Framework         & Tensorflow Lite\footnote{https://www.tensorflow.org/lite}     & Edge           \\ \hline
Communication                       & MQTT\footnote{https://mqtt.org/}                & Edge/Fog       \\ \hline
Dispatching                         & Node-RED\footnote{https://nodered.org/}            & Fog            \\ \hline
Telemetries Data Storage            & InfluxDB\footnote{https://www.influxdata.com/}            & Cloud          \\ \hline
Data Visualization                  & Grafana\footnote{https://grafana.com/}             & Cloud          \\ \hline
Virtualization through DT & Eclipse Ditto\footnote{https://www.eclipse.org/ditto/}       & Cloud          \\ \hline
\end{tabular}%
}
\end{table}

The proposed TMS employs a network of distributed software modules that operate across the system architecture, as detailed in Table \ref{tab:purposes}. To adhere to EI principles, deep learning algorithms are deployed in proximity to data sources. Additionally, the USB or Coral Camera, Environmental Sensor Board, Gas Sensor Board, and ESP32 must be located at the network edge to obtain real-time telemetry directly from the field. Once the data has been collected and processed, it is transmitted to the core module of the system via an MQTT Broker, which was implemented using Eclipse Mosquitto. The Node-RED platform was utilized to manage the backend logic of the system, including the creation of a wiring connection between hardware devices and software modules (dispatching function), as well as the establishment of rules, thresholds, and conditions. The Node-RED module has the capability to (i) retrieve or update DTs' status information via the REST API of Eclipse Ditto; (ii) maintain data persistence using two NoSQL databases, namely MongoDB for DTs status and InfluxDB for telemetries; and (iii) facilitate data visualization through the use of Grafana's dashboards. Both MQTT Broker and the Node-RED application module resides on the fog layer, while Eclipse Ditto, MongoDB, InfluxDB and Grafana services, can be hosted on a cloud instance.
\section{Preliminary Results and Future Development}
\label{sec:results}
By harnessing the concepts expounded in the preceding sections, it is theoretically apparent that implementing EI confers several advantages over cloud-based alternatives. These benefits encompass reduced latency, lower bandwidth consumption, improved energy efficiency, and heightened privacy. Our approach centers on integrating EI technology, allowing us to process data locally and in proximity to its origin. This eliminates the requirement to transfer sensitive data to remote servers, ensuring its security. To quantify the efficacy of our approach, we conducted a comparative analysis of our TMS performance using two different configurations (shown in Table \ref{tab:hw_edge_cloud}) for the vehicle detection task:
\begin{itemize}
    \item The \textbf{EI-based solution}, which we propose, involves carrying out the vehicle detection on the Google Coral Dev Board equipped with the Google Edge TPU. The inferences' output is then transmitted to other nodes in the system's architecture.
    \item The \textbf{Cloud-centric solution} involves capturing real-time video frames of the road at the edge and transmitting them to a standard cloud instance, such as the Intel\textsuperscript{®} Core™ i7-6820HQ, 2.70GHz x 8 for the vehicle detection task. After processing, the inference results are then transmitted back to other nodes in the system's architecture.
\end{itemize}
\begin{table}[htbp]
  \caption{Hardware configuration: Edge- vs  Cloud-based deployment}
  \label{tab:hw_edge_cloud}
  \begin{center}
    \begin{tabular}{|c|c|c|}
        \hline
        & \textbf{Edge} & \textbf{Cloud} \\
        \hline
        %\textbf{Model} & Google Coral Dev & HP ZBook 17 G3\\
        %& Board &\\
        %\hline
        \textbf{Processor} & Google Edge TPU & Intel\textsuperscript{®} Core™ i7-6820HQ, \\
        & coprocessor & 2.70GHz x 8\\
        \hline
        \textbf{Memory (RAM)} & 4 GB LPDDR4 & 32 GB\\
        \hline
        \textbf{Graphics} & Integrated GC7000 & AMD® Bonaire/ Mesa \\
        & Lite Graphics & Intel® HD Graphics 530\\
        \hline
    \end{tabular}
  \end{center}
\end{table}
%We have hence evaluated different deep learning models and different system configurations according to typical performance metrics like model accuracy, inference latency and generated data traffic.

We have conducted a comprehensive evaluation encompassing a range of deep learning models and system configurations, while giving due consideration to key performance metrics such as model accuracy, inference latency, and generated data traffic. Initially, we assessed several models designed for vehicle detection, with a focus on their compatibility with an environment based on Edge TPU, thereby harnessing the capabilities of this hardware platform. Our evaluation encompassed a total of five models, comprising three pre-trained models (namely, SSD MobileNet V1, SSD MobileNet V2, and SSDLite MobileDet) sourced from Google Coral Documentation, as well as two models trained using the Transfer Learning technique on the EfficientDet architecture. Specifically, these two models are referred to as the MTD Model, trained on the Mini Traffic Detection (MTD) dataset, and the TI Model, trained on the Traffic Images (TI) dataset. To evaluate the performance of each model, the \textit{mean average precision} (\textit{mAP}) was measured, which is the primary metric used in Common Object in COntext (COCO) evaluation metrics. 

\begin{figure}[htbp]
    \centering
    \includegraphics[width=0.48\textwidth]{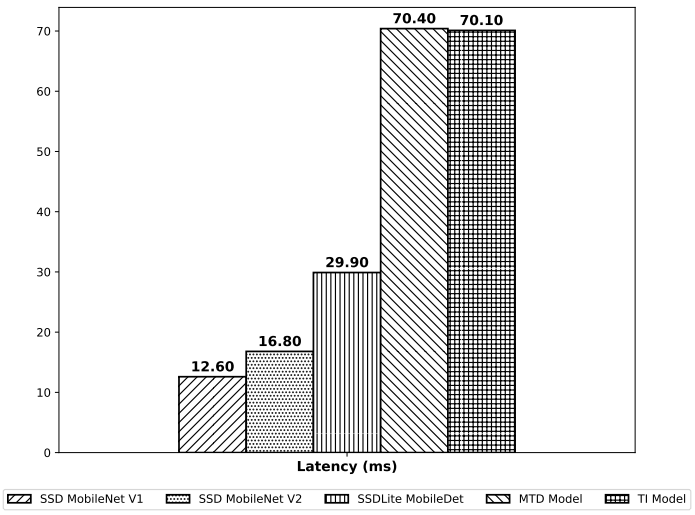}
    \caption{\label{fig:latency} Pre-trained and Re-trained models: latency per inference}
\end{figure}

\begin{figure}[htbp]
    \centering
        \includegraphics[width=0.48\textwidth]{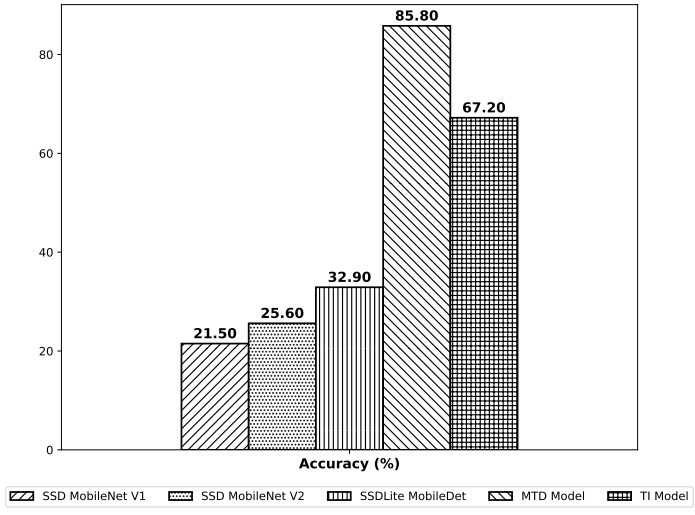}
    \caption{\label{fig:accuracy} Pre-trained and Re-trained models: mean average precision}
\end{figure}

Figure \ref{fig:latency} and Figure \ref{fig:accuracy} demonstrate that the pre-trained models perform exceptionally well in terms of latency per inference, with none of the three exceeding 30ms per frame. Among the tested models, the fastest one is the SSD MobileNet V1, with a latency of just 12.6ms, as confirmed by our tests. However, while these models excel in latency, they fall short in accuracy, struggling to detect certain classes of objects and occasionally failing to detect objects even when they are present in the frame. Differently, the re-trained MTD Model and TI Model offer higher precision at the cost of lower responsiveness, with both models displaying a latency of around 70ms per frame, which is more than twice as long as the lowest performing pre-trained model. Despite this, the MTD model achieves the highest accuracy among the tested models, with a \textit{mAP} score of 85.8\%. Based on our evaluation, we have chosen to use the MTD model in our proposed EI-based TMS. This decision was driven by the MTD model's good accuracy and its reasonable latency per inference, which strike an ideal balance for our application's requirements. The Edge TPU's faster processing time offers several advantages, including lower power consumption and memory savings. Slower models typically require more power to complete tasks, but the Edge TPU's optimized architecture reduces processing time and thus power consumption.

Having identified a suitable model for the vehicle detection task, we started evaluating the edge- and cloud-based deployments for our TMS. With respect to generated data traffic, at the edge, the camera module captures video frames using a Python algorithm, with each frame having a file size of around \textit{1,936 Mb}. This results in significant data traffic that can cause network latency of up to \textit{43ms} per frame, even when using the Coral Dev Board's 802.11ac interface at 2.4GHz for wider coverage and a realistic speed of up to 450Mbps. Furthermore, this approach increases power dissipation in the edge device and has a significant impact on bandwidth consumption. In our experiments, we found that a one-day interaction between the edge device and cloud server, with a \textit{20fps} (frames per second) bit rate, resulted in the transmission of \textit{1,729,000 frames}. This equates to \textit{3.34Tb} of data moving over the network per day. However, processing the frame locally and transmitting only the inference results to a remote server requires an MQTT message of \textit{67 bytes}, which includes the ID of the packet, the payload, and the header for control purposes. This approach results in a reduction of data traffic by a factor of 10,000 (only \textit{0.872Gb}) compared to the previous case.

%\begin{table}[htbp]
%  \caption{Performance Evaluation: Edge- vs  Cloud-based deployment}
%  \label{tab:edge&cloud}
%  \begin{center}
%    \begin{tabular}{|c|c|c|}
%        \hline
%        & \textbf{Edge} & \textbf{Cloud} \\
%        \hline
%        \textbf{Model Latency (per inference)} & 70.40ms & 112.32ms\\
%        \hline
%        \textbf{Network Latency (per frame)} & 0 & 34ms\\
%        \hline
%        \textbf{Data Traffic (per frame)} & 0.54Kb & 1.936Mb\\
%        \hline
%        \textbf{Data Traffic (per day)} &0.872Gb & 3.34Tb\\
%        \hline
%    \end{tabular}
%  \end{center}
%\end{table}

%\textcolor{red}{The results presented in Figure \ref{fig:edgeVScloud} demonstrate that, on average, the Edge-based configuration outperforms the Cloud-based configuration by a factor of 2.2 in terms of latency per inference. Among the tested models, the MTD model achieved the highest accuracy (see Figure \ref{fig:accuracy}), with a \textit{mean average precision} (\textit{mAP}) score of 85.8\%. Despite having a latency per inference that is more than twice as long as the lowest performing pre-trained model, the MTD model offers the best trade-off between accuracy and latency. Therefore, we have chosen to use the MTD model in the proposed EI-based TMS.}

\subsection{Future Developments}
Our current solution primarily addresses the EI aspects of a TMS, specifically the vehicle detection task. However, leveraging on the potentials of ODT and on Eclipse Ditto (provides excellent support for DT technology, enabling us to manage and update DT data), we plan to enhance the synthetic sensing capabilities of our TMS to better detect and respond to emergent situations. This includes developing models that, exploiting the existing general-purpose devices, can identify unusual patterns in traffic flow or environmental conditions, such as sudden changes in temperature or humidity, which could indicate an emerging situation. In addition to this, we aim to develop other event-based rules that can help automate responses to different scenarios. For instance, we plan to implement a system that sends alarms to city governance if pollution levels exceed a certain threshold, triggering an emergency response plan. We also plan to improve the noise sensors to detect the presence of noisy vehicles, which can help regulate traffic lights in the noisy lanes and could be useful in identifying emergency vehicles. Finally, we aim to enhance our analysis by leveraging real-world data to evaluate the effectiveness of our TMS. This evaluation will enable the identification of areas for improvement, including unanticipated situations such as adverse weather conditions. Factors like fog or strong winds can profoundly influence vehicle detection and pollutant collection. By taking these unexpected scenarios into account, we can fine-tune our system accordingly, ensuring its optimal performance.
\section{Conclusion}
\label{sec:conclusion}
The digitalization process has revolutionized the way cities operate by enhancing data ubiquity and automation. The synergy of EI, novel sensing techniques and DT has the potential to enhance the overall efficiency of cities, creating smarter and more sustainable solutions. With this aim in mind, we outlined an innovative TMS for smart cities which leverages, exactly, on EI and ODT for data gathering and local processing. Our system architecture consists of hardware and software components optimized for the efficient collection, processing, and utilization of data. Additionally, real-time inference information and the current counter are sent to the fog node, which runs the backend logic and updates the DT information. By adopting an EI-based approach and pushing intelligence as close as possible to data sources, we can reduce inference latencies by 40\%, eliminating network latency for the vehicle detection task, and decreasing data traffic by a factor of 10,000 per day compared to the cloud-centric approach. The strategy implemented in our proposal holds huge potential to revolutionize how we sense and equip our cities, presenting an efficient and cost-effective means of gathering valuable data where cutting-edge technologies work together towards creating a smarter, greener, and more livable city. We strongly believe that the ODT approach offers a notable opportunity for improving the performance of large-scale living systems, extending beyond the scope of smart cities. These systems can include, but are not limited to, areas such as smart agriculture (which can benefit from improved harvest quality), energy management (where efficiency and sustainability can be enhanced), building automation (by improving occupant comfort), manufacturing (for especially for predictive maintenance), and environmental monitoring (to facilitate the tracking and prediction of changes in natural systems such as weather patterns and water quality).

%By combining existing infrastructure with advanced sensing capabilities, we can significantly enhance the performance of autonomous vehicles and other technologies, leading to better traffic management and reduced emissions. The method proposed in our study emphasizes the need for practicality and sustainability in solution development, preventing the squandering of valuable resources and effort in the pursuit of innovation. Overall, our research offers a promising path towards shaping the future of urban living, where cutting-edge technologies work together towards creating a smarter, greener, and more livable city.

\bibliographystyle{IEEEtran}
\bibliography{IEEEabrv,bibliography}

\end{document}